# Theoretical restrictions on longest implicit timescales in Markov state models of biomolecular dynamics

*Anton V. Sinitskiy*[1], *Vijay S. Pande*[1,2]

[1] Department of Chemistry, [2] Department of Structural Biology and Department of Computer Science, Stanford University, Stanford, CA 94305, USA. Correspondence should be addressed to A.V.S (sinitskiy@stanford.edu) or V.S.P. (pande@stanford.edu).

Markov state models (MSMs) have been widely used to analyze computer simulations of various biomolecular systems. They can capture conformational transitions much slower than an average or maximal length of a single molecular dynamics (MD) trajectory from the set of trajectories used to build the MSM. A rule of thumb claiming that the slowest implicit timescale captured by an MSM should be comparable by the order of magnitude to the aggregate duration of all MD trajectories used to build this MSM has been known in the field. However, this rule have never been formally proved. In this work, we present analytical results for the slowest timescale in several types of MSMs, supporting the above rule. We conclude that the slowest implicit timescale equals the product of the aggregate sampling and four factors that quantify: (1) how much statistics on the conformational transitions corresponding to the longest implicit timescale is available, (2) how good the sampling of the destination Markov state is, (3) the gain in statistics from using a sliding window for counting transitions between Markov states, and (4) a bias in the estimate of the implicit timescale arising from finite sampling of the conformational transitions. We demonstrate that in many practically important cases all these four factors are on the order of unity, and we analyze possible scenarios that could lead to their significant deviation from unity. Overall, we provide for the first time analytical results on the slowest timescales captured by MSMs. These results can guide further practical applications of MSMs to biomolecular dynamics and allow for higher computational efficiency of simulations.

# I. INTRODUCTION

Markov state models (MSMs) have been one of the main tools for computational studies of biomolecules.[1-3] These models represent the dynamics of a biomolecule in the form of discrete transitions between a finite number of states. In MSMs, the system is assumed to be memoryless, that is, the probabilities of transition between Markov states depend only on the current state, but not on the previous states of the system.

MSMs are particularly useful when multiple molecular dynamics (MD) trajectories are available.[4] In this case, an MSM of a simulated biomolecular system can integrate all the dynamic information from different trajectories, yielding estimates of the equilibrium and long-term dynamic properties of a biomolecular system under investigation. It has been known for a long time that this approach can capture slow dynamics of the system, with characteristic timescales much longer than lengths of single MD trajectories in the set of trajectories employed to construct an MSM.

How far in timescale can MSMs proceed in aggregating data from multiple trajectories? To the best of our knowledge, no strict answer to this question based on analytical derivations have ever been published. However, an empirical rule has been known in the community of specialists applying MSMs to study biomolecules, namely, that the slowest implicit timescale captured by an MSM can go up to the values on the same order of magnitude as the aggregate sampling achieved in the used MD dataset. In previous studies of specific biomolecular systems with MSMs based on MD simulations, reported longest implicit timescales turned out to be smaller than the aggregate sampling by at least one order of magnitude[5-17] or comparable to the aggregate sampling.[18-22]

In this paper, we analyze several MSMs with different topologies to uncover the reasons for the above empirical rule. These MSMs represent typical cases when slow dynamics is observed in biomolecular systems.

# II. THEORY

## A. Simplest two-state MSM of a rare conformational transition

A simplest two-state Markov model with rare transitions between two states (Fig. 1) can capture many aspects of the slowest-timescale dynamics of larger MSMs, while allowing for exact analytical solutions. State 1 in this model represents the native state of the biomolecule (corresponding to most Markov states in a multistate MSM with frequent transitions between them), while state 0 represents its metastable state (corresponding to an outlying Markov state(s) in a multistate MSM reached in only a small fraction of available MD trajectories).

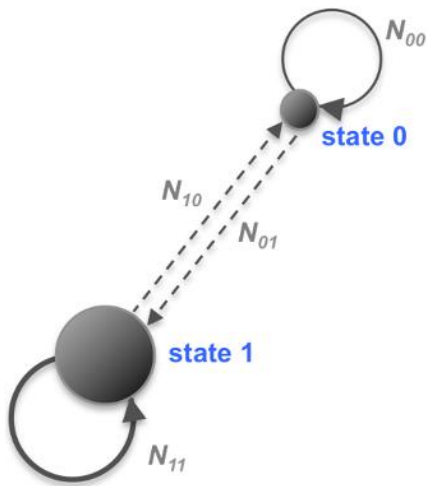

**Fig. 1.** Two-state MSM allows for exact analytical solutions, and can also serve as a simplified version of a multistate MSM with rare transitions between two subgraphs.

For a given set of MD trajectories, the count matrix $C$ equals

$$C = \begin{pmatrix} N_{00} & N_{01} \\ N_{10} & N - N_{00} - N_{01} - N_{10} \end{pmatrix}, \quad (1)$$

where $N$ is the total number of observed transitions, $N_{00}$, $N_{01}$ and $N_{10}$ are the numbers of transitions from state 0 to state 0, from state 0 to state 1 and from state 1 to state 0, respectively. Then, the transition matrix $T$ can be found as follows (the direct symmetrization of the number of counts is used to enforce the reversibility of the transition matrix):

$$T = \begin{pmatrix} 1 - \dfrac{\check{S}}{\check{S}_0 + \check{S}} & \dfrac{\check{S}}{\check{S}_0 + \check{S}} \\ \dfrac{\check{S}}{1 - \check{S}_0 - \check{S}} & 1 - \dfrac{\check{S}}{1 - \check{S}_0 - \check{S}} \end{pmatrix}, \quad (2)$$

where $\check{S}_0 = N_{00}/N$ is the relative sampling of state 0, and $\check{S} = (N_{01}+N_{10})/2N$ is the relative sampling of transitions between states 0 and 1. Due to the physical interpretation of the model,

$$\check{S} \ll \check{S}_0 < 1. \quad (3)$$

One of the eigenvalues of matrix $T$ is $\lambda_0 = 1$ (corresponding to the stationary/equilibrium state), while the other is

$$\lambda_1 = 1 - \dfrac{\check{S}}{(\check{S}_0 + \check{S})(1 - \check{S}_0 - \check{S})}. \quad (4)$$

The corresponding implicit timescale $\tau_1$, in the units of the lagtime $\tau_{lagtime}$, is

$$\dfrac{\tau_1}{\tau_{lagtime}} = -\dfrac{1}{\ln\left[1 - \dfrac{\check{S}}{(\check{S}_0 + \check{S})(1 - \check{S}_0 - \check{S})}\right]}, \quad (5)$$

which, taking into consideration Eq. (3), simplifies to

$$\dfrac{\tau_1}{\tau_{lagtime}} \approx \dfrac{2\check{S}_0(1 - \check{S}_0)N}{N_{01} + N_{10}}. \quad (6)$$

It follows from Eq. (6) that the longest implicit timescale computed from an MSM can be written as the product

$$\tau_1 = f_{rare\ state} f_{transitions} f_{sw} T, \quad (7)$$

where $T$ is the total sampling (in the units of time), and $f_{rare\ state}$, $f_{transitions}$ and $f_{sw}$ are dimensionless factors. Specifically, $f_{rare\ state}$ is determined by the fraction of MD frames in which the system is in the rare state (state 0):

$$f_{rare\ state} = 2\check{S}_0(1-\check{S}_0), \tag{8}$$

$f_{transitions}$ depends on how much sampling of the rare transition events is available:

$$f_{transitions} = \frac{1}{N_{01}+N_{10}}, \tag{9}$$

and $f_{sw}$ is determined by the ratio of the lagtime to the output frequency in the used MD trajectories, and also may include a correction for the effect of counting transitions using a sliding window. The derivation above refers to the case where a sliding window is not used, and the intermediate points in the MD trajectories are ignored, then

$$f_{sw} = \frac{\ddagger_{lagtime}}{T/N}. \tag{10}$$

In practice, either the lagtime is taken to be equal to the output frequency, implying $f_{sw}=1$, or the lagtime is greater than the output frequency, and a sliding window is used. In the latter case, the transitions cannot be considered as statistically independent, and Eq. (10), corresponding to independent data, presents an overestimation. Therefore, in the case of a sliding window, we can give an estimate of

$$1 \leq f_{sw} \leq \frac{\ddagger_{lagtime}}{T/N}. \tag{11}$$

Currently, we do not have an analytical expression for $f_{sw}$ in the case of a sliding window transition counts. We speculate that in this case a reasonable estimate might be

$$f_{sw} \sim \sqrt{\frac{\ddagger_{lagtime}}{T/N}}. \tag{12}$$

Detailed analysis of factors $f_{rare\ state}$, $f_{transitions}$ and $f_{sw}$, as well as the corresponding implications for using MSMs to study the dynamics of biomolecules, are provided below, in the Discussion section.

Next, consider how an estimate of the longest timescale from a finite sample relates to the exact value of this timescale. Denote the exact probabilities of transition from state 0 to state 1 and from state 1 to state 0 per lagtime by $\Omega_{01}$ and $\Omega_{10}$, respectively. These variables are similar to the above variables $\omega_{01}$ and $\omega_{10}$, with the difference that $\Omega_{01}$ and $\Omega_{10}$ are the exact probabilities, while $\omega_{01}$ and $\omega_{10}$ are the corresponding estimates from a finite statistical sample (a given set of MD trajectories). We do not introduce a separate variable for the exact probability of staying in state 0, because due to the physical interpretation of the model, there are much more events of staying in state 0 than transitioning between states 0 and 1, and therefore, the difference between the exact probability of staying in state 0 and its estimate from a finite sample can be neglected in comparison to the differences between $\Omega_{01}$ and $\omega_{01}$, and between $\Omega_{10}$ and $\omega_{10}$.

By analogy with Eq. (6), the exact implicit timescale, in the units of the lagtime, is given by:

$$\frac{\ddagger_1^{exact}}{\ddagger_{lagtime}} = \frac{2\check{S}_0(1-\check{S}_0)}{\Omega_{01}+\Omega_{10}}. \tag{13}$$

On the other hand, the expectation value of the implicit timescale estimated from a finite sample is given by:

$$\frac{\langle\ddagger_1\rangle}{\ddagger_{lagtime}} = 2\check{S}_0(1-\check{S}_0)N\left\langle\frac{1}{N_{01}+N_{10}}\right\rangle, \tag{14}$$

where the values of $N_{01}$ and $N_{10}$ are random variables following the Poisson distribution:

$$N_{01} \sim Pois(\Omega_{01}N), \qquad N_{10} \sim Pois(\Omega_{10}N), \tag{15}$$

and the average on the right hand side of Eq. (14) is taken over the zero-truncated distribution (elementary events with $N_{01}=0$ and simultaneously $N_{10}=0$ are omitted, because the corresponding MSMs do not provide a finite estimate of the implicit timescale). Therefore,

$$\frac{\langle\ddagger_1\rangle}{\ddagger_{lagtime}} = 2\check{S}_0(1-\check{S}_0)N\frac{\displaystyle\sum_{\substack{N_{01},N_{10}=0\\\neg(N_{01}=0\wedge N_{10}=0)}}^{\infty}\frac{1}{N_{01}+N_{10}}\frac{(\Omega_{01}N)^{N_{01}}}{N_{01}!}\frac{(\Omega_{10}N)^{N_{10}}}{N_{10}!}e^{-(\Omega_{01}+\Omega_{10})N}}{\displaystyle\sum_{\substack{N_{01},N_{10}=0\\\neg(N_{01}=0\wedge N_{10}=0)}}^{\infty}\frac{(\Omega_{01}N)^{N_{01}}}{N_{01}!}\frac{(\Omega_{10}N)^{N_{10}}}{N_{10}!}e^{-(\Omega_{01}+\Omega_{10})N}}. \tag{16}$$

The sums in the right hand side of Eq. (16) can be computed by using the following integral representation:

$$\frac{1}{N_{01}+N_{10}} = \int_0^1 \frac{d\xi}{\xi} \xi^{N_{01}+N_{10}}, \quad (17)$$

switching the order of integration and summation in Eq. (16), and recognizing the Taylor series of the exponential function. As a result, the ratio of the average estimated timescale to its true value can be simplified to:

$$\frac{\langle t_1 \rangle}{t_1^{exact}} = \frac{ne^{-n}\left[Ei(n) - \chi - \ln(n)\right]}{1 - e^{-n}}, \quad (18)$$

where $n = (\pi_{01} + \pi_{10})N$ is the average expected number of transitions between states 0 and 1, $Ei$ is the exponential integral function, and $\chi$ is the Euler–Mascheroni constant. Detailed analysis of the function on the right hand side of Eq. (18) is provided in the Discussion section.

To sum up, the simplest two-state MSM with rare transitions reveals many important aspects of the relationship between the slowest timescale estimated from finite data, the exact value of this timescale, and the aggregate sampling. But how transferable these results are to MSMs with more states? We explore this question in the next three sections (Sec. II.B-D).

**B. Perturbative analysis of an arbitrarily large MSM with a rarely sampled state**

Consider an MSM with $N_B$ closely connected states and one more state with rare transitions to/from it (Fig. 2; the notation "B" for the set of closely connected states is used for comparability of the formulas below with the analysis in Sec. II.C). Without loss of generality, we denote this rarely sampled state as state 0, and all other states as states 1, …, $N_B$.

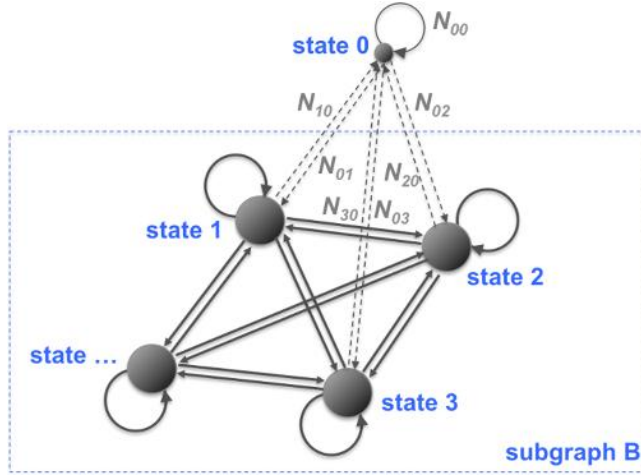

**Fig. 2.** Multistate MSM with rare transitions between one state (state 0) and all other states is a typical situation where long implicit timescales emerge in practice.

The count matrix can be written as

$$C = \begin{pmatrix} N_{00} & N_{01} & N_{02} & \cdots & N_{0N_B} \\ N_{10} & N_{11} & N_{12} & \cdots & N_{1N_B} \\ N_{20} & N_{21} & N_{22} & & \\ \vdots & & & \ddots & \\ N_{N_B 0} & N_{N_B 1} & & & N_{N_B N_B} \end{pmatrix}, \qquad (19)$$

where $N_{ij}$ is the number of observed transitions from state $i$ to state $j$. Then, the transition matrix $T$ can be written in a block-diagonal form as follows (the direct symmetrization of the number of counts is implied):

$$T = \begin{pmatrix} 1 - \dfrac{\check{S}}{\check{S}_0 + \check{S}} & T^{0B} \\ T^{B0} & T^{BB} \end{pmatrix}, \qquad (20)$$

where $T^{B0}$ is a column formed by the elements

$$T_i^{B0} = \frac{(\check{S}_{0i} + \check{S}_{i0})/2}{(\check{S}_{0i} + \check{S}_{i0})/2 + \frac{1}{N}\sum_{j=1}^{N_{st}} \frac{N_{ij} + N_{ji}}{2}}, \qquad i = 1,...,N_B, \tag{21}$$

$T^{0B}$ is a row formed by the elements

$$T_i^{0B} = \frac{(\check{S}_{0i} + \check{S}_{i0})/2}{\check{S}_0 + \check{S}}, \qquad i = 1,...,N_B, \tag{22}$$

$T^{BB}$ is an $N_B \times N_B$ square matrix formed by the elements

$$T_{ij}^{BB} = \frac{(N_{ij} + N_{ji})/2}{N(\check{S}_{0i} + \check{S}_{i0})/2 + \sum_{k=1}^{N_{st}}(N_{ik} + N_{ki})/2}, \qquad i = 1,...,N_B, \quad j = 1,...,N_B, \tag{23}$$

$š_0 = N_{00}/N$ is the relative sampling of state 0, $š_{ij} = N_{ij}/N$, and $\overline{w}$ is the relative sampling of transitions to or from state 0:

$$\check{S} = \sum_{j=1}^{N_{st}} \frac{\check{S}_{0i} + \check{S}_{i0}}{2}. \tag{24}$$

If transitions to and from state 0 are relatively rare:

$$\check{S}_{0i} \ll 1, \quad \check{S}_{i0} \ll 1, \qquad i = 1,...,N_B, \tag{25}$$

then eigenvalues of the transition matrix $T$, and therefore the implicit timescales, can be found by perturbation expansion. First, define the zeroth-order terms. In the limit of

$$\check{S}_{0i} \to 0, \quad \check{S}_{i0} \to 0, \qquad i = 1,...,N_B, \tag{26}$$

the transition matrix reduces to

$$T \to \tilde{T} = \begin{pmatrix} 1 & 0 & \cdots & 0 \\ 0 & & & \\ \vdots & & \tilde{T}^{BB} & \\ 0 & & & \end{pmatrix}, \tag{27}$$

where $\tilde{T}^B$ is an $N_B \times N_B$ square matrix formed by the elements

$$\tilde{T}^{BB}_{ij} = \frac{N_{ij} + N_{ji}}{\sum_{k=1}^{N_{st}} N_{ik} + N_{ki}}, \qquad i=1,...,N_B, \quad j=1,...,N_B. \tag{28}$$

Matrix $\tilde{T}$ has two eigenvalues equal to 1, one of which corresponds to the stationary state in the subsystem of states 1, ..., $N_B$, and the other to the absence of transitions (infinitely large characteristic timescale) between state 0 and the other states. The corresponding left and right eigenvectors are:

$$v_0^\dagger = \begin{pmatrix} 0 & \tilde{p}_1 & \cdots & \tilde{p}_{N_B} \end{pmatrix}, \quad w_0 = \begin{pmatrix} 0 \\ 1 \\ \vdots \\ 1 \end{pmatrix}, \quad v_1^\dagger = \begin{pmatrix} 1 & 0 & \cdots & 0 \end{pmatrix}, \quad w_1 = \begin{pmatrix} 1 \\ 0 \\ \vdots \\ 0 \end{pmatrix}, \tag{29}$$

where $\tilde{p}_1, \ldots, \tilde{p}_{N_B}$ are the occupations of states 1, ..., $N_B$ in subgraph B at equilibrium in the absence of transitions between state 0 and subgraph B. The eigenvectors satisfy the conditions

$$\begin{aligned}
\tilde{T} w_0 = w_0, \quad \tilde{T} w_1 = w_1, \quad v_0^\dagger \tilde{T} = v_0^\dagger, \quad v_1^\dagger \tilde{T} = v_1^\dagger, \\
v_0^\dagger w_0 = 1, \quad v_1^\dagger w_1 = 1, \quad v_0^\dagger w_1 = 0, \quad v_1^\dagger w_0 = 0,
\end{aligned} \tag{30}$$

All other eigenvalues of matrix $\tilde{T}$ are less than 1, corresponding to finite implicit timescales of transitions in the subsystem of states 1, ..., $N_B$ (we assumed that this subsystem is fully connected). Perturbations to the first two eigenvalues can be found as the eigenvalues of the matrix

$$\begin{pmatrix} v_0^\dagger (T-\tilde{T}) w_0 & v_0^\dagger (T-\tilde{T}) w_1 \\ v_1^\dagger (T-\tilde{T}) w_0 & v_1^\dagger (T-\tilde{T}) w_1 \end{pmatrix} = \begin{pmatrix} -\dfrac{\check{S}}{\check{S}_0} & \dfrac{\check{S}}{\check{S}_0} \\ \dfrac{\check{S}}{1-\check{S}_0} & -\dfrac{\check{S}}{1-\check{S}_0} \end{pmatrix} \tag{31}$$

[the right hand side contains the leading terms computed with the use of Eq. (25)]. The eigenvalues of this matrix are:

$$u\}_0 = 0, \quad u\}_1 = -\frac{\check{S}}{\check{S}_0(1-\check{S}_0)}. \tag{32}$$

As expected, one of these corrections is zero, since the perturbed matrix still must have one eigenvalue exactly equal to 1, corresponding to the stationary state of the whole system. The other correction is nonzero, leading to a finite implicit timescale:

$$\frac{\ddagger_1}{\ddagger_{lagtime}} = \frac{2\check{S}_0(1-\check{S}_0)N}{\sum_{i=1}^{N_{st}}(N_{0i}+N_{i0})}. \tag{33}$$

(again, an approximation based on Eq. (25) was used). This result is strikingly simple and similar to Eq. (6) derived for a two-state MSM. The only difference is that instead of the total number of transitions between states 0 and 1, this expression contains the total number of transitions to or from state 0. No properties of the subsystem of states 1, …, $N_B$, such as the equilibrium populations of each of these states or probabilities of transitions between them, are left in the final result, Eq. (33), even though they were present in the intermediate formulas. Consequently, the general relationship between the slowest implicit timescale and the total sampling, Eq. (7), holds in this case, with the only change that the factor $f_{transitions}$ now assumes the following form:

$$f_{transitions} = \frac{1}{\sum_{i=1}^{N_B}(N_{0i}+N_{i0})}. \tag{34}$$

Moreover, the analysis of the relationship between the exact value of the implicit timescale and its estimates from finite data can be directly extended to this case, too. The expectation value of the timescale can be expressed as

$$\frac{\langle\ddagger_1\rangle}{\ddagger_{lagtime}} = 2\check{S}_0(1-\check{S}_0)N\left\langle\frac{1}{\sum_{i=1}^{N_B}(N_{0i}+N_{i0})}\right\rangle, \tag{35}$$

where the expectation value on the right hand side of Eq. (35) is taken over zero-truncated Poisson distributions of the variables $N_{0i}$ and $N_{i0}$,

$$N_{0i} \sim Pois(\Omega_{0i}N), \qquad N_{i0} \sim Pois(\Omega_{i0}N), \qquad i=1,\ldots,N_B, \qquad (36)$$

where $\Omega_{0i}$ and $\Omega_{i0}$ are the exact probabilities of transitions from state 0 to state $i$ or vice versa per lagtime. Then, Eq. (35) can be rewritten as

$$\frac{\langle \ddagger_1 \rangle}{\ddagger_{lagtime}} = 2\check{S}_0(1-\check{S}_0)N \times$$

$$\times \frac{\sum\limits_{\substack{N_{0i},N_{i0}=0 \\ \neg \forall i(N_{0i}0 \wedge N_{i0}=0)}}^{\infty} \frac{1}{\sum\limits_{i=1}^{N_B}(N_{0i}+N_{i0})} \prod\limits_{i=1}^{N_B} \frac{(\Omega_{0i}N)^{N_{0i}}}{N_{0i}!} \frac{(\Omega_{i0}N)^{N_{i0}}}{N_{i0}!} e^{-\sum\limits_{i=1}^{N_B}(\Omega_{0i}+\Omega_{i0})N}}{\sum\limits_{\substack{N_{0i},N_{i0}=0 \\ \neg \forall i(N_{0i}0 \wedge N_{i0}=0)}}^{\infty} \prod\limits_{i=1}^{N_B} \frac{(\Omega_{0i}N)^{N_{0i}}}{N_{0i}!} \frac{(\Omega_{i0}N)^{N_{i0}}}{N_{i0}!} e^{-\sum\limits_{i=1}^{N_B}(\Omega_{0i}+\Omega_{i0})N}}. \qquad (37)$$

Computing the sums leads to the result given by Eq. (18), with the following definition of $n$ as the expectation value of the total number of transitions to or from state 0:

$$n = N \sum_{i=1}^{N_B}(\Omega_{0i}+\Omega_{i0}). \qquad (38)$$

Thus, the analysis of the bias in the estimated slowest implicit timescale introduced by a finite sampling is also transferable from the simplest two-state MSM to an MSM with multiple states, if transitions to and from one of these states are rarely sampled.

### C. Perturbative analysis of a large MSM with rare transitions between two subgraphs

Another typical case leading to large implicit timescales is the existence of two (or more) subgraphs with rare transitions between them (Fig. 3). Perturbative analysis of this model can be performed similar to the analysis presented in Sec. II.B.

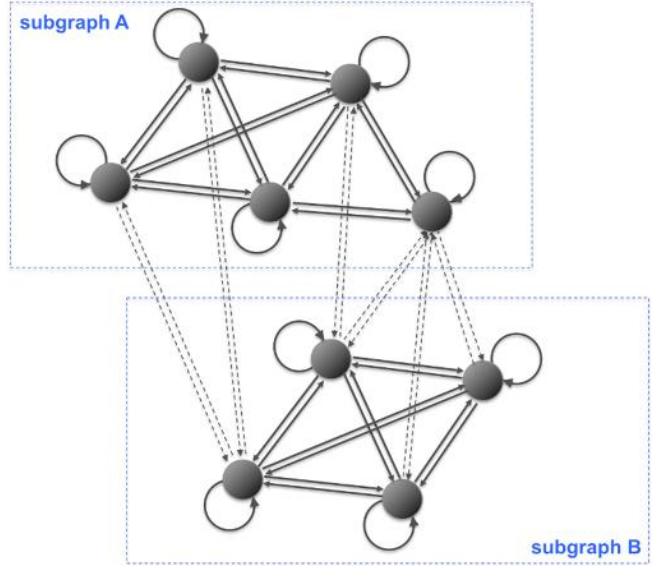

**Fig. 3.** Multistate MSM with rare transitions between two subgraphs is another typical situation where long implicit timescales emerge in practice.

The count matrix can be written in a block-diagonal form:

$$C = \begin{pmatrix} C^{AA} & C^{AB} \\ C^{BA} & C^{BB} \end{pmatrix}, \quad (39)$$

where matrices $C^{AA}$, $C^{AB}$, $C^{BA}$ and $C^{BB}$ contain the number of transition between different states within subgraph A, from states in subgraph A to states in subgraph B, from states in subgraph B to states in subgraph A, and between states within subgraph B, respectively. Then, the transition matrix also assumes the block-diagonal form:

$$T = \begin{pmatrix} T^{AA} & T^{AB} \\ T^{BA} & T^{BB} \end{pmatrix}, \quad (40)$$

As the unperturbed state, we choose the MSM without transitions between subgraphs A and B:

$$\tilde{T} = \begin{pmatrix} \tilde{T}^{AA} & 0 \\ 0 & \tilde{T}^{BB} \end{pmatrix}, \quad (41)$$

where

$$\tilde{T}^{AA} = \lim_{C^{AB}, C^{BA} \to 0} T^{AA}, \qquad \tilde{T}^{BB} = \lim_{C^{AB}, C^{BA} \to 0} T^{BB}, \qquad (42)$$

If each of subgraphs A and B is connected, then the matrix $\tilde{T}$ has exactly two eigenvalues equal to 1, corresponding to the stationary states of isolated subgraphs A and B, while all other eigenvalues of $\tilde{T}$ are less than 1. The left and right eigenvectors corresponding to the first two eigenvalues are:

$$v_0^\dagger = \begin{pmatrix} \tilde{p}^A & \vec{0}^\dagger \end{pmatrix}, \quad w_0 = \begin{pmatrix} \vec{1} \\ \vec{0} \end{pmatrix}, \quad v_1^\dagger = \begin{pmatrix} \vec{0}^\dagger & \tilde{p}^B \end{pmatrix}, \quad w_1 = \begin{pmatrix} \vec{0} \\ \vec{1} \end{pmatrix}, \qquad (43)$$

where $\vec{0}$ and $\vec{1}$ are columns with all elements equal to 0 or 1, respectively, and $\tilde{p}^A$ and $\tilde{p}^B$ are stationary (equilibrium) distributions for subgraphs A and B in the absence of transitions between these two subgraphs:

$$\tilde{p}^A \tilde{T}_{AA} = \tilde{p}^A, \qquad \tilde{p}^B \tilde{T}_{BB} = \tilde{p}^B. \qquad (44)$$

Then, perturbations to the first two eigenvalues can be found as the eigenvalues of the matrix

$$\begin{pmatrix} v_0^\dagger (T-\tilde{T}) w_0 & v_0^\dagger (T-\tilde{T}) w_1 \\ v_1^\dagger (T-\tilde{T}) w_0 & v_1^\dagger (T-\tilde{T}) w_1 \end{pmatrix} = \begin{pmatrix} -\dfrac{N_{AB}+N_{BA}}{2N_{AA}} & \dfrac{N_{AB}+N_{BA}}{2N_{AA}} \\ \dfrac{N_{AB}+N_{BA}}{2N_{BB}} & -\dfrac{N_{AB}+N_{BA}}{2N_{BB}} \end{pmatrix}, \qquad (45)$$

where $N_{AA}$, $N_{AB}$, $N_{BA}$ and $N_{BB}$ are the total number of transitions between different states within subgraph A, from states in subgraph A to states in subgraph B, from states in subgraph B to states in subgraph A, and between states within subgraph B, respectively:

$$N_{AA} = \sum_{i_A=1}^{N_{st,A}} \sum_{j_A=1}^{N_{st,A}} C_{i_A j_A}^{AA}, \quad N_{AB} = \sum_{i_A=1}^{N_{st,A}} \sum_{j_B=1}^{N_{st,B}} C_{i_A j_B}^{AB}, \quad N_{BA} = \sum_{i_B=1}^{N_{st,B}} \sum_{j_A=1}^{N_{st,A}} C_{i_B j_A}^{BA}, \quad N_{BB} = \sum_{i_B=1}^{N_{st,B}} \sum_{j_B=1}^{N_{st,B}} C_{i_B j_B}^{BB}. \qquad (46)$$

For example,

$$v_0^\dagger \left(T - \tilde{T}\right) w_0 = \sum_{i_A=1}^{N_{st,A}} \sum_{j_A=1}^{N_{st,A}} \tilde{p}_{i_A}^A \left(T_{i_A j_A}^{AA} - \tilde{T}_{i_A j_A}^{AA}\right)$$

$$= \sum_{i_A=1}^{N_{st,A}} \sum_{j_A=1}^{N_{st,A}} \tilde{p}_{i_A}^A \left( \frac{\frac{C_{i_A j_A}^{AA} + C_{j_A i_A}^{AA}}{2}}{\sum_{k_A=1}^{N_{st,A}} \frac{C_{i_A k_A}^{AA} + C_{k_A i_A}^{AA}}{2} + \sum_{k_B=1}^{N_{st,B}} \frac{C_{i_A k_B}^{AB} + C_{k_B i_A}^{BA}}{2}} - \frac{\frac{C_{i_A j_A}^{AA} + C_{j_A i_A}^{AA}}{2}}{\sum_{k_A=1}^{N_{st,A}} \frac{C_{i_A k_A}^{AA} + C_{k_A i_A}^{AA}}{2}} \right) \quad (47)$$

$$= -\sum_{i_A=1}^{N_{st,A}} \tilde{p}_{i_A}^A \frac{\sum_{k_B=1}^{N_{st,B}} \frac{C_{i_A k_B}^{AB} + C_{k_B i_A}^{BA}}{2}}{\sum_{k_A=1}^{N_{st,A}} \frac{C_{i_A k_A}^{AA} + C_{k_A i_A}^{AA}}{2} + \sum_{k_B=1}^{N_{st,B}} \frac{C_{i_A k_B}^{AB} + C_{k_B i_A}^{BA}}{2}}.$$

If transitions between the two subgraphs are rare in comparison to transitions within each subgraph, the leading term in Eq. (47) is

$$v_0^\dagger \left(T - \tilde{T}\right) w_0 \approx -\sum_{i_A=1}^{N_{st,A}} \tilde{p}_{i_A}^A \frac{\sum_{k_B=1}^{N_{st,B}} \frac{C_{i_A k_B}^{AB} + C_{k_B i_A}^{BA}}{2}}{\sum_{k_A=1}^{N_{st,A}} \frac{C_{i_A k_A}^{AA} + C_{k_A i_A}^{AA}}{2}} = -\sum_{i_A=1}^{N_{st,A}} \tilde{p}_{i_A}^A \frac{\sum_{k_B=1}^{N_{st,B}} \frac{C_{i_A k_B}^{AB} + C_{k_B i_A}^{BA}}{2}}{\tilde{p}_{i_A}^A \sum_{k_A=1}^{N_{st,A}} \sum_{l_A=1}^{N_{st,A}} \frac{C_{l_A k_A}^{AA} + C_{k_A l_A}^{AA}}{2}}$$

$$= -\frac{\sum_{i_A=1}^{N_{st,A}} \sum_{k_B=1}^{N_{st,B}} \frac{C_{i_A k_B}^{AB} + C_{k_B i_A}^{BA}}{2}}{\sum_{k_A=1}^{N_{st,A}} \sum_{l_A=1}^{N_{st,A}} \frac{C_{l_A k_A}^{AA} + C_{k_A l_A}^{AA}}{2}} = -\frac{N_{AB} + N_{BA}}{2N_{AA}}. \quad (48)$$

The other elements of the matrix in Eq. (45) can be computed in a similar way. Then, the eigenvalues of this matrix are:

$$u\}_0 = 0, \quad u\}_1 = -\frac{(N_{AB} + N_{BA})(N_{AA} + N_{BB})}{2N_{AA} N_{BB}} \approx -\frac{(N_{AB} + N_{BA})N}{2N_{AA} N_{BB}}, \quad (49)$$

where $N$ is the total number of observed transitions:

$$N = N_{AA} + N_{BB} + N_{AB} + N_{BA} \approx N_{AA} + N_{BB}. \quad (50)$$

As expected, one of the eigenvalues stays unchanged and equal to 1, corresponding to the stationary state of the whole system. The correction to the other eigenvalue is nonzero, leading to a finite implicit timescale:

$$\frac{\ddagger_1}{\ddagger_{lagtime}} = \frac{2\check{S}_A \check{S}_B N}{N_{AB} + N_{BA}}, \qquad (51)$$

where $\check{S}_A = N_{AA}/N$, $\check{S}_B = N_{BB}/N$. This result is also similar to the expression for the implicit timescale derived from the simplest two-state MSM, Eq. (6). The subsequent analysis, including the relationship between the slowest implicit timescale and the total sampling, Eq. (7), holds in this case, too, with the factor $f_{transitions}$ now assuming the following form:

$$f_{transitions} = \frac{1}{N_{AB} + N_{BA}}. \qquad (52)$$

The analysis of the relationship between the exact value of the implicit timescale and its estimates from finite datasets can also be directly extended to this case, leading to the result given in Eq. (18), with the following definition of $n$ as the expectation value of the total number of transitions between subgraphs A and B:

$$n = N \left( \sum_{i_A=1}^{N_{st,A}} \sum_{j_B=1}^{N_{st,B}} \Omega_{i_A j_B}^{AB} + \sum_{i_B=1}^{N_{st,B}} \sum_{j_A=1}^{N_{st,A}} \Omega_{i_B j_A}^{BA} \right), \qquad (53)$$

where $\Omega$ values are the exact probabilities of transitions between specific states in subgraphs A and B.

**D. Three-state MSM with rare transitions to two states**

The analysis above in Sec. II.A-C referred to MSMs with one slow implicit timescale. In this section, we analyze cooperative effects between two different sets of rare transitions, leading to the appearance of two slow implicit timescales. Taking into consideration the results of Sec. II.B-C, where we demonstrated that the relationships for the slow implicit timescale are transferrable from the simplest two-state MSM to more general models, in this section we limit ourselves to the analysis of a three-state MSM, assuming that the results will be qualitatively applicable to more general models with two slow implicit timescales and an arbitrary number of Markov states.

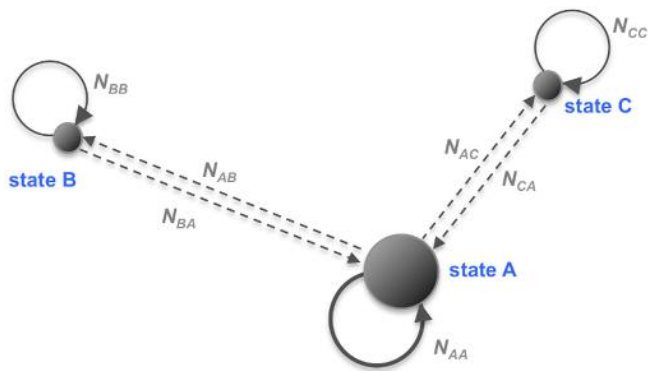

**Fig. 4.** Three-state MSM with rare transitions between state A and state B, and state A and state C can serve as a simplest example allowing for an exact analysis of coupling of two processes with slow timescales.

This model (Fig. 4) includes three states: state A, corresponding to the native geometry of a biomolecule, and states B and C, which correspond to two metastable states of the biomolecule. Transitions between A and B, and between A and C are rare, and there are no direct transitions between B and C. The count matrix can be written as

$$C = \begin{pmatrix} N_{AA} & N_{AB} & N_{AC} \\ N_{BA} & N_{BB} & 0 \\ N_{CA} & 0 & N_{CC} \end{pmatrix}, \quad (54)$$

where $N_{ij}$ is the number of observed transitions from state $i$ to state $j$. The transition matrix $T$ (after a direct symmetrization of the number of counts) assumes the following form:

$$T = \begin{pmatrix} 1-r_{AB}-r_{AC} & r_{AB} & r_{AC} \\ r_{BA} & 1-r_{BA} & 0 \\ r_{CA} & 0 & 1-r_{CA} \end{pmatrix}, \quad (55)$$

where

$$r_{AB} = \frac{N_{AB} + N_{BA}}{2N_{AA} + N_{AB} + N_{BA} + N_{AC} + N_{CA}}, \quad r_{AC} = \frac{N_{AC} + N_{CA}}{2N_{AA} + N_{AB} + N_{BA} + N_{AC} + N_{CA}},$$
$$r_{AB} = \frac{N_{AB} + N_{BA}}{2N_{BB} + N_{AB} + N_{BA}}, \quad r_{AC} = \frac{N_{AC} + N_{CA}}{2N_{CC} + N_{AC} + N_{CA}}. \tag{56}$$

The transition matrix has the following eigenvalues:

$$\lambda_0 = 1, \quad \lambda_{1,2} = 1 - \frac{(r_{AB} + r_{BA} + r_{AC} + r_{CA}) \pm \sqrt{(r_{AB} + r_{BA} - r_{AC} - r_{CA})^2 + 4r_{AB}r_{AC}}}{2}, \tag{57}$$

and hence, the exact expressions for the two implicit timescales are

$$\frac{t_{1,2}}{t_{lagtime}} = -\frac{1}{\ln\left[1 - \frac{(r_{AB} + r_{BA} + r_{AC} + r_{CA}) \pm \sqrt{(r_{AB} + r_{BA} - r_{AC} - r_{CA})^2 + 4r_{AB}r_{AC}}}{2}\right]}. \tag{58}$$

Unlike previous cases, it does not seem possible to analytically simplify Eq. (58) for small $N_{AB}/N$, $N_{BA}/N$, $N_{AC}/N$, $N_{CA}/N$ and finite $\pi_B = N_{BB}/N$ and $\pi_C = N_{CC}/N$. However, in the practically important case of

$$\frac{N_{AB}}{N}, \frac{N_{BA}}{N}, \frac{N_{AC}}{N}, \frac{N_{CA}}{N} \ll \check{\pi}_B, \check{\pi}_C \ll 1, \tag{59}$$

Eq. (58) simplifies to

$$\frac{t_1}{t_{lagtime}} = \frac{2\check{\pi}_B(1-\check{\pi}_B)N}{N_{AB} + N_{BA}}\left(1 + O(\check{\pi}^2)\right),$$
$$\frac{t_2}{t_{lagtime}} = \frac{2\check{\pi}_C(1-\check{\pi}_C)N}{N_{AC} + N_{CA}}\left(1 + O(\check{\pi}^2)\right), \tag{60}$$

where the small parameter $\check{\pi} = \max(\check{\pi}_B, \check{\pi}_C)$. These expressions for the implicit timescales are similar to the result for the two-state MSM, Eq. (6), and demonstrate decoupling of rare transitions from each other, at least in the case of rare sampling of the corresponding metastable states.

The analysis of the bias in the estimates of implicit timescales can be performed similar to the cases analyzed in Sec. II.A-C, leading to the results similar to Eq. (18), namely:

$$\frac{\langle \ddagger_1 \rangle}{\ddagger_1^{exact}} = \frac{n_1 e^{-n_1} \left[ Ei(n_1) - \chi - \ln(n_1) \right]}{1 - e^{-n_1}}, \quad n_1 = N(\Omega_{AB} + \Omega_{BA}),$$
$$\frac{\langle \ddagger_2 \rangle}{\ddagger_2^{exact}} = \frac{n_2 e^{-n_2} \left[ Ei(n_2) - \chi - \ln(n_2) \right]}{1 - e^{-n_2}}, \quad n_2 = N(\Omega_{AC} + \Omega_{CA}), \tag{61}$$

where values are the exact probabilities of transitions between the corresponding states.

## III. DISCUSSION

In all four models analyzed above, the slowest implicit timescale(s) of rarely sampled conformational transitions have been expressed as a product of the aggregate sampling $T$ and three dimensionless factors:

$$\ddagger = f_{rare\ state} f_{transitions} f_{sw} T, \tag{62}$$

Moreover, it has been demonstrated that this estimate from a finite MD dataset is, in general, biased, while the unbiased estimate is given by

$$\ddagger^{exact} = f_{rare\ state} f_{transitions} f_{sw} f_{bias} T. \tag{63}$$

Consider now what might be typical values of these four dimensionless factors and their product in most cases, and what scenario may lead to significantly greater or smaller values of this product.

The first factor $f_{rare\ state}$ is determined by the shares of MD frames in which the system is in one or the other subgraph. Its maximum value is 1/2, which is achieved when half of all sampled frames refer to one subgraph, and the other half to the other subgraph. If the metastable state is less sampled, which is a typical situation, then $f_{rare\ state}$ is approximately equal to the fraction of frames in this metastable state ($\Omega_0$ in Sec. II.A-B, $\Omega_A$ or $\Omega_B$ in Sec. II.C, or $\Omega_B$ or $\Omega_C$ in Sec. II.D). In such cases, the longest timescale is limited by the time the system spent in the rare state, rather than the aggregate sampling for all possible states.

The maximum value of $f_{transitions}$ is 1, and it is achieved when only one transition between two subgraphs occurred in the simulations. This factor falls to 1/2 if two transitions are observed in either direction, or one transition in each direction. With an increase in the sampling, $f_{transitions}$ decreases, but not so drastically, and remains on the order of unity as long as the total number of observed rare transitions stays a single-digit number.

The third factor $f_{sw}$ is greater than 1 if a sliding window is used for counting transitions, and equals 1 otherwise. In practice, in most situations the lagtime is ~10-100 times greater than the output frequency, a sliding window is used, and therefore, based on the assumption of Eq. (12), $f_{sw}$ ~ 3-10.

Finally, the bias $f_{bias}$ in the timescale estimate, according to eqs. (18) and (61), can be written as

$$f_{bias} = \frac{1-e^{-n}}{ne^{-n}\left[Ei(n) - \text{x} - \ln(n)\right]}, \quad (64)$$

The right hand side of Eq. (64) monotonically decreases from $+\infty$ to 0.7574 when $n$ goes from 0 to 3.75, and after than monotonically increases, with the asymptotic value of 1 reached in the limit of $n \to \infty$ (Fig. 5). Therefore, the estimate of the implicit timescale is unbiased not only in the limit of infinite sampling, due to the law of large numbers, but also at $n$ = 1.503, due to a cancellation of the effects of zero-truncation in the Poisson distributions on the numerator and denominator in eqs. (16), (37) and similar equations for MSMs with more states. In the practically important range of $n$ ~ 1–10, the bias may be positive (at $n$ < 1.503) or negative ($n$ > 1.503), reaching +30% at $n$ = 1 and –24% at $n$ = 3.75. This analysis of the systematic bias is based on the approximation of rare transitions between subgraphs in various types of MSMs [eqs. (3), (25), or (59)], which allowed us to proceed with analytical derivations beyond the results previously published in the literature,[23-26] and obtain the analytical result given by Eq. (64).

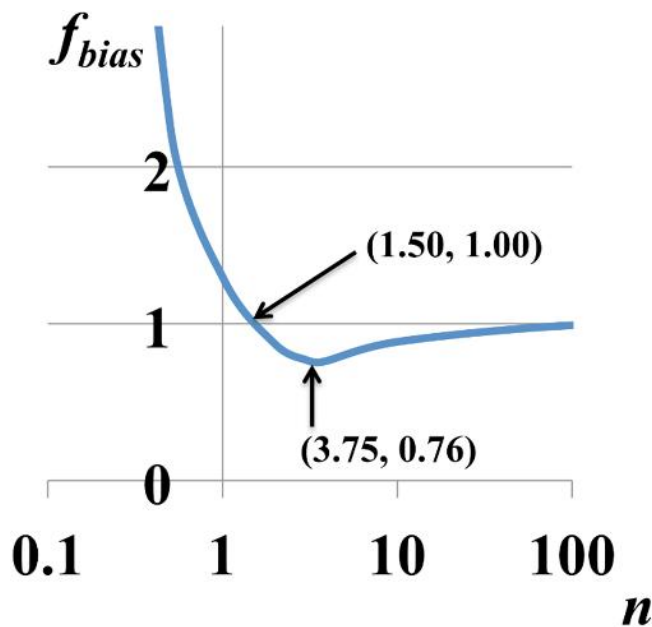

**Fig. 5.** The dependence of the bias $f_{bias}$ on the expected number of rare transitions $n$ has a nontrivial character, as follows from Eq. (64).

Overall, the product of factors $f_{rare\ state}$, $f_{transitions}$, $f_{sw}$ and, if the unbiased estimate of the timescale is required, $f_{bias}$, in many practical situations may be ~1, plus/minus one order of magnitude, and hence, the longest implicit timescale estimated from an MSM is indeed comparable by the order of magnitude to the total sampling used to build this MSM. In principle, the longest timescale can exceed the aggregate sampling, with $f_{sw} > 1$ as the necessary condition required for it. Given the total sampling, longer timescales can be reached when *the rarely visited subgraph is more sampled*, when *transitions for which the implicit timescale is sought for are less* sampled (which, however, undermines the statistical robustness of the estimated timescale), and when *a sliding window is used* for counting transitions. Reseeding additional MD simulations from rarely visited state(s) has a twofold effect on the achievable timescales. On the one hand, it increases the fraction of frames referring to the rarely sampled state(s), thereby increasing $f_{rare\ state}$. On the other hand, it increases the number of transitions to and from rarely sampled state(s), thereby decreasing $f_{transitions}$. The net effect should be an increase in the achievable timescales, because such a reseeding proportionally increases the fraction of the frames in the rare states and the number of transitions from the rare states, but does not affect that much the number of transitions from the native state to the rare states.

## IV. CONCLUSIONS

In this work, we have analytically investigated several representatives MSMs with topologies favoring the appearance of slow implicit timescales. In all cases, unexpectedly simple relationships between the slowest implicit timescale and the aggregate sampling were derived. Our analytical results confirm the empirical rule that the slowest implicit timescale can reach the same order of magnitude as the aggregate sampling, and reveal the quantitative effects of sampling rare states and transitions to or from them on the achievable timescales. This work can serve as a basis for a rational design of MD simulation campaigns, allowing for modeling longer-timescale conformational changes in biomolecular systems at lower cost, that is with less computer sampling.

**ACKNOWLEDGEMENTS**

The authors were funded by NIH R01-GM062868.